\documentclass[final,11pt,3p,sort&compress,twocolumn]{elsarticle} 
\makeatletter
\def\ps@pprintTitle{%
 \let\@oddhead\@empty
 \let\@evenhead\@empty
 \def\@oddfoot{\reset@font\hfil\thepage\hfil}%
 \let\@evenfoot\@oddfoot}

\biboptions{super} 

\usepackage[latin2]{inputenc}
\usepackage{amssymb}
\usepackage{amsthm}
\usepackage{amsmath}
\usepackage{graphicx,cancel}
\usepackage{ulem}
\usepackage{color}
\usepackage{fixltx2e} 
\usepackage{balance} 
\usepackage{float}	
\usepackage{dblfloatfix} 
\usepackage{booktabs}
\usepackage[size=small, labelfont=bf]{caption} 
\usepackage{hyperref}
\hypersetup{
    colorlinks=true, 
    linkcolor=blue,  
    citecolor=blue,  
}
    
\renewcommand{\_}[1]{{}_{\mathrm{#1}}}

\begin{document}

\begin{frontmatter}

\title{\LARGE{\textbf{Investigation of ultra-thin layers of bis(phthalocyaninato)lutetium(III) on graphite}}}


\author[]{\large{Lars~Smykalla}}
\author[]{Pavel~Shukrynau}
\author[]{Michael~Hietschold}
\address{Technische Universität Chemnitz, Institute of Physics, Solid Surfaces Analysis Group,\\ D-09107 Chemnitz, Germany}

\begin{abstract}
We present a comprehensive study of the adsorption of bis(phthalocyaninato)lutetium(III) (LuPc$\_2$) on highly oriented pyrolytic graphite(0001) (HOPG). The growth and self-assembly of the molecular layers as well as the electronic structure has been investigated systematically using scanning tunnelling microscopy and scanning tunnelling spectroscopy combined with density functional theory (DFT) calculations and molecular mechanics simulations. We reveal that the adsorption of LuPc$\_2$ leads to the formation of a square-like close-packed structure on the almost inert surface of HOPG, which is corroborated by simulations. Moreover, we observed a parallel orientation of the LuPc$\_2$ molecules in the first monolayer, whereas in subsequent layers an increasing tilt out of the surface plane was found. Tip$-$sample distance-dependent tunnelling spectroscopy measurements allowed us to detect a shift in the energy positions of the peaks assigned to the lowest unoccupied molecular orbital toward the Fermi energy with decreasing tip$-$sample separation.
\end{abstract}

\end{frontmatter}

\section{Introduction}

Nanostructures formed through the self-organization of organic molecules on crystalline substrates gained tremendous interest over the past decade. Thin organic films can modify surfaces to make them suitable for many different applications reaching from simple electronic interconnections and sensors\cite{Rodriguez-Mendez1999} to sophisticated spintronic devices\cite{Bogani2008}. This leads to the need to investigate the molecular arrangement and electronic and magnetic properties of organic molecules adsorbed on a substrate to gain detailed insight into the interactions between the molecules within a layer as well as between the molecules and the surface. From this point of view, metallo\-phthalo\-cyanines have been intensively studied because of their interesting electronic and magnetic properties and their high chemical and thermal stability.\cite{Torre2007} Recently, rare-earth double-decker phthalo\-cyanine molecules moved into focus\cite{Komeda2011,Zhang2009a,Vitali2008,Gomez-Segura2006,Takami2006,Takami2010,Biagi2010} after it was found that they behave as single-molecule magnets and show a long magnetisation relaxation time.\cite{Ishikawa2003,Ishikawa2004,Branzoli2010} Their ability to store the information of a spin state can be used in future spintronic applications.\cite{Ishikawa2005} Bisphthalocyanine complexes have a one-electron ligand-oxidised ``sandwich'' structure with a rare-earth metal coordinated between two phthalocyanines (Pc), as shown in Figure~\ref{fig:subML}a. In the molecule, the unpaired electron is delocalised over the $\pi$ orbitals of both phthalocyanine ligands.\cite{VanCott1995,Ishikawa2004a} Substitution of the hydrogen atoms of the Pc ligand by other functional groups can modify and, in some way, tune intermolecular bonding\cite{Ye2006,Takami2007} as well as the bonding strength to the substrate\cite{Kyatskaya2009}. All of this implies the possibility to get different structures and, as a consequence, to improve transport properties for sophisticated devices. Previous investigations on dinuclear lanthanide triple-decker molecules\cite{Guyon1998} have shown the magnetic interaction between the ions\cite{Ishikawa2002} and the orientation-dependent ordering of different modified triple-decker complexes\cite{Lei2008}. Furthermore, for bis(phthalocyaninato)\-lutetium(III) (LuPc$\_2$) on Ag(111), the tunnelling transport revealed a negative differential resistance in the negative bias voltage range.\cite{Toader2011b} The neutral LuPc$\_2$ has a molecular spin of 1/2 and is the most conductive among the lanthanide Pc$\_2$ complexes with an electrical bulk conductivity even higher by more than six orders of magnitude compared with other metallo\-phthalocyanines.\cite{Andre1985,Jones1997} However, still very little is known about the adsorption of LuPc$\_2$ on surfaces. The atomic geometry and the electronic structure of the molecule in the layers as well as the bonding between the molecules and the substrate are the main problems, which have to be understood. In this work, we investigated in detail structural and electronic properties of the LuPc$\_2$ / HOPG system in a wide range of coverage, starting from the very early stages of molecular adsorption up to the growth of layers with the height of a few monolayers (MLs).

\section{Methods}
\subsection{Experimental details}

Substrates were prepared by cleaving highly oriented pyrolytic graphite(0001) using adhesive tape and subsequently annealing in ultrahigh vacuum (UHV) at 600$\,^{\circ}$C. LuPc$\_2$ powder was filled into a Knudsen cell and purified at a temperature slightly below the sublimation temperature in UHV for several days. Ultrathin layers of LuPc$\_2$ were then deposited by organic molecular beam epitaxy at 340$\,^{\circ}$C for submonolayer coverage and 400$\,^{\circ}$C for multilayer coverages on the cleaned substrate, held at room temperature. The molecular layers were slightly post-annealed at a temperature of 100$\,^{\circ}$C for several times over the course of the experiments to allow the formation of large highly ordered molecular islands. The scanning tunnelling microscopy (STM) experiments were performed with a variable-temperature STM (Omicron) under UHV conditions. The base pressure in the UHV chamber was in the range of $10^{-10}\,$mbar. The sample was cooled to $\approx 30\,$K using a helium flow cryostat to freeze the motion of the molecules on the weakly interacting graphite substrate to allow stable and highly resolved imaging and measurements. Tungsten tips for the STM were electrochemically etched and annealed in UHV by bringing them close to a heated tungsten wire to remove the tungsten oxide layer. All STM images were measured in constant current mode with a tunnelling current of 100\,pA. The indicated bias voltages were applied to the sample. STM data was processed with the WSxM\cite{Horcas2007} software, whereby moderate low pass filtering was applied for noise reduction. Tip$-$sample distance-dependent tunnelling spectroscopy was performed by measuring the current$-$voltage-dependence over a single molecule in the ordered layer for increasing set point tunnelling currents and a fixed bias voltage before opening the feedback loop, thus leading to a decreasing tip$-$sample distance in subsequent characteristics. $(\text{d}I/\text{d}V)/(I/V)$, which is proportional to the local density of states,\cite{Feenstra1987} was then calculated from the average $I(V)$ curve from at least 100 single spectra at a fixed set point.

\subsection{Computational details}

Calculations were carried out with the Gaussian 03 package.\cite{Gaussian03} Thereby, for the density functional theory (DFT)
calculations, the unrestricted mPW1PW91\cite{Adamo1998a} functional and the SDD\cite{Fuentealba1982,Cao2001} basis set were used to optimise the geometry of the free molecule and to calculate the single-electron energies and wave functions of the Kohn-Sham orbitals. For molecular mechanics simulations, the Universal force field\cite{Rappi1992} (UFF) was applied to calculate potential energy surfaces for an arrangement of four LuPc$\_2$ molecules. Force fields are suitable to describe a van der Waals system\cite{Gopakumar2008} like the physisorption of a layer of LuPc$\_2$ on graphite and the intermolecular interaction between the molecules. The calculated total energy is the sum of all noncovalent energies emerging of the Lennard-Jones potential and the electrostatic interaction of the partial charges between each atom. The binding energy is calculated by subtracting the total energy of the arrangement from the sum of the energies of the isolated parts. The four LuPc$\_2$ molecules are placed on the corners of a parallelogram to model the molecular lattice, whereby the orientation of the Pc ligands is the same for all molecules. This model is sufficient to give qualitative results for the intermolecular interactions for one molecule with its surrounding molecules in the molecular layer. To simulate the structure of the ML on HOPG, the molecular arrangement is placed on a supporting graphene layer, which is used as a model to include the influence of the graphite substrate in the calculation. Thereby, the molecules lie flat on the supporting surface with one molecular lattice vector parallel to one lattice vector of graphene. No substrate is used for the simulation of the tilt angle out of plane of the molecules in the top molecular layer of a multilayer because otherwise it would restrict the molecular tilt, as phthalocyanines adsorb with the molecular plane parallel to the surface in the first layer.

\section{Results and Discussion}

\begin{figure}[tb]
\centering
\includegraphics[width=0.49\textwidth]{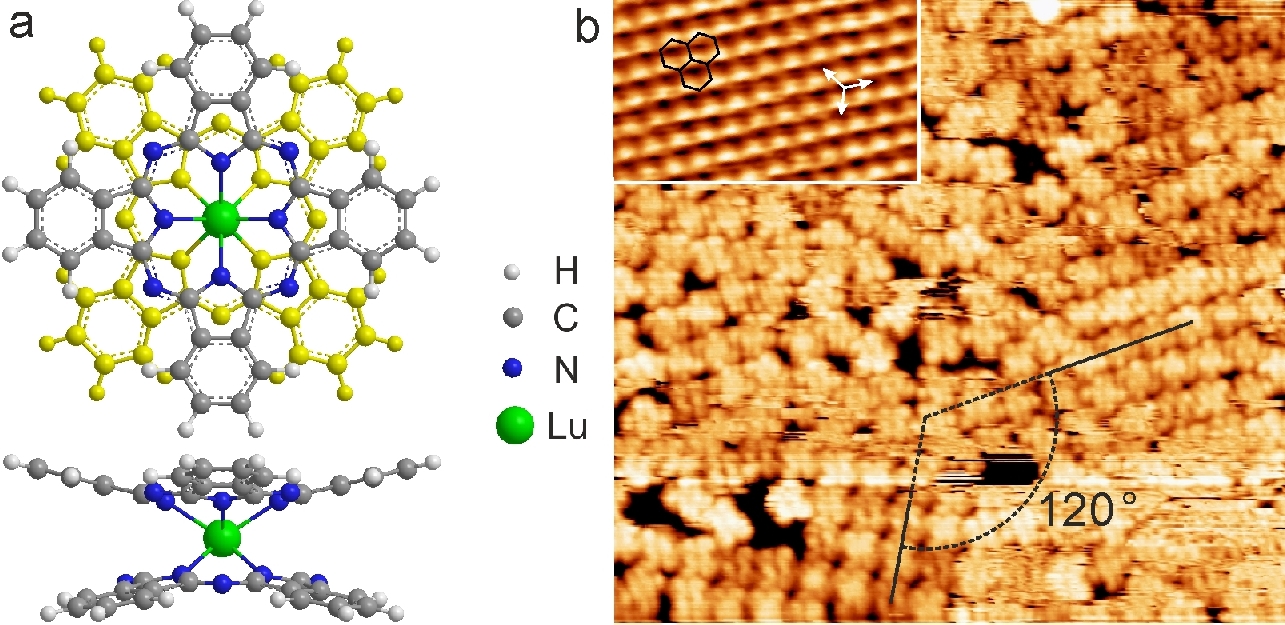}
\caption{(a) Geometry of a DFT-optimised LuPc$\_2$ molecule. (b) STM image ($26.6\times 24.8$\,nm$^2$, $V = -1.5\,$V) of a molecular island of LuPc$\_2$ on HOPG with both disordered and ordered areas at submonolayer coverage. An atomic resolution image ($3.0\times 1.8\,$nm$^2$) of the HOPG substrate underneath the molecular layer is inserted in panel b.}
\label{fig:subML}
\hrulefill
\end{figure}

\begin{figure*}[bt]
\centering
\includegraphics[width=1.0\textwidth]{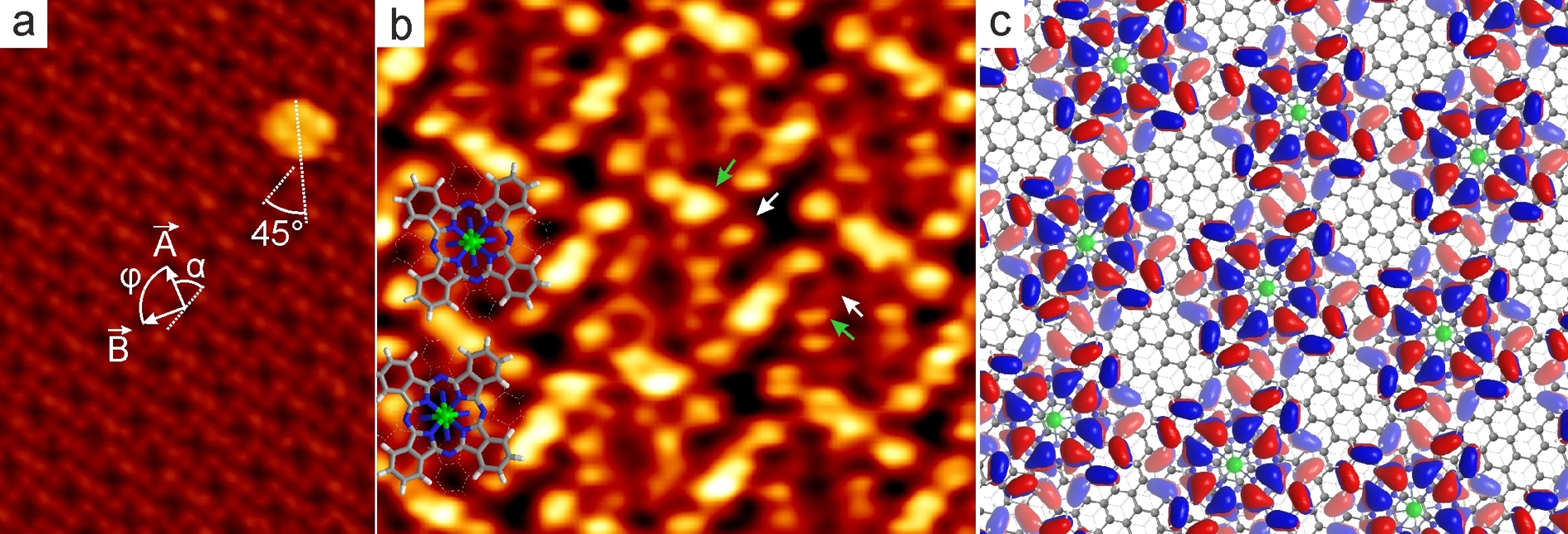}
\caption{(a) STM image ($12.5\times 17.7\,$nm$^2$) of the first layer of LuPc$\_2$ adsorbed on HOPG imaged with $V = +1.2\,V$. (b) Highly resolved STM image ($4.8\times 4.3\,$nm$^2$) of a monolayer of LuPc$\_2$ showing intramolecular features of the molecular appearance at negative bias voltage of $V = -1.5\,V$. The model of the upper Pc ligand is superimposed on the STM image. The green and white arrows indicate the small height differences between neighbouring protrusions of high electron density. (c) Model for the epitaxy of LuPc$\_2$ on HOPG. For comparison for each molecule the calculated shape of the singly occupied molecular orbital is shown.}
\label{fig:struc}
\hrulefill
\end{figure*}

Room-temperature deposition of LuPc$\_2$ onto the HOPG substrate at the coverage of 1/4 of a closed ML leads to the formation of the large, partially ordered molecular islands of variable shape and size, as can be seen in Figure~\ref{fig:subML}b. Despite the fact that the measurement was done at a temperature of 30\,K, where the mobility of the molecules is strongly reduced, the rearrangement of molecules and increase of ordered regions during scanning was still observed. This can be explained by the weaker interaction of graphite with adsorbed molecules compared with metal substrates\cite{Toader2011b} and by the additional influence of the scanning tip. After the samples were slightly annealed, most molecules are arranged in a highly ordered structure. By comparison with the underlying HOPG (see the inset in Figure~\ref{fig:subML}b), it was found that the structure adsorbs with a preferential direction of one lattice vector parallel to a lattice vector of graphite. Consequently, between the corresponding lattice vectors of different ordered domains, an angle of $\approx 120^{\circ}$ is measured. In Figure~\ref{fig:struc}, the typical geometry of the molecular layer is presented.
The lengths of the lattice vectors of the ordered ML were measured to be $A = (1.48\pm 0.03)\,\mathrm{nm}$ and $B = (1.54\pm 0.03)\,\mathrm{nm}$ with an angle between them of $\varphi = (88\pm 1)^{\circ}$ and the angle of an axis through the upper Pc ligand with vector $\vec{A}$ of $a = (66\pm 2)^{\circ}$. This results in the epitaxial relation

$$\begin{pmatrix} \vec{A}\\ \vec{B} \end{pmatrix} = \begin{bmatrix} 6.0 & 0\\ -3.4 & 7.2 \end{bmatrix} \begin{pmatrix} \vec{a_1}\\ \vec{a_2} \end{pmatrix}$$
\begin{figure*}[t]
\centering
\includegraphics[width=1.0\textwidth]{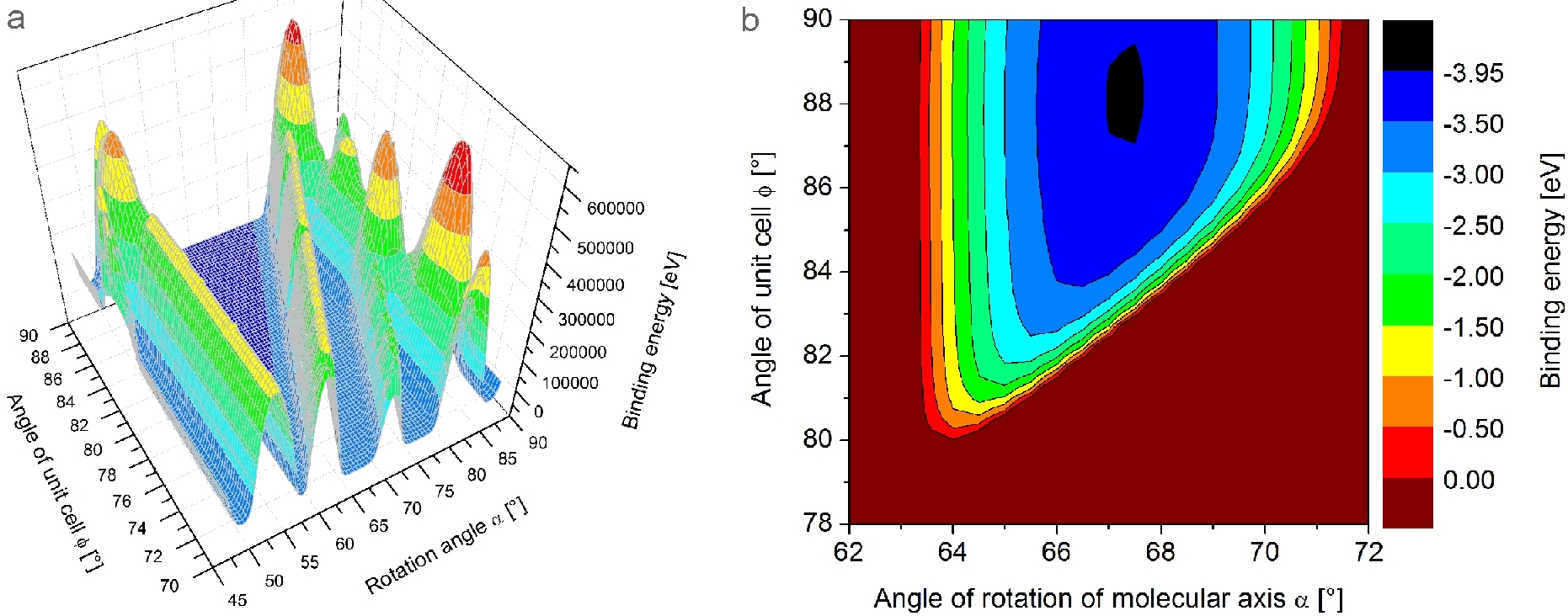}
\caption{(a) 3D plot of the potential energy surface of four LuPc$\_2$ at the corners of an unit cell by variation of the rotation angle of the molecules and the angle of the unit cell at fixed intermolecular distances. A graphene sheet is included in this calculation to simulate the influence of the substrate on the molecular arrangement. (b) Magnification of the potential energy surface around the lowest minimum.}
\label{fig:LuPc2_uff}
\hrulefill
\end{figure*}

between the lattice vectors $\vec{A}$ and $\vec{B}$ of the molecular layer and $\vec{a_1}$ and $\vec{a_2}$ of the graphite substrate. The islands are of ML thickness and often have individual LuPc$\_2$ molecules adsorbed on top of them (Figure~\ref{fig:struc}a). The molecules have a height of circa 0.35\,nm, which shows that we are dealing with double-decker Pc and not single Pc molecules, which have a typical apparent height of circa 0.15 nm and would occur if LuPc$\_2$ would have been decomposed. The LuPc$\_2$ molecule has a cross-like appearance, which is typical for phthalocyanines, with a depression in the centre, hence neither the bottom Pc ligand nor the lutetium atom has a significant contribution to the STM image. In Figure~\ref{fig:struc}a, one molecule in the layer is imaged with one protrusion over each of the four phenylene groups of the upper Pc ligand. The highly resolved STM image (as presented in Figure~\ref{fig:struc}b) reveals the internal electronic structure of the molecule. One can see that the single LuPc$\_2$ molecule consists of eight protrusions from the electronic $\pi$ system distributed over the porphyrazin ring with a depression in the center and four pairs of protrusions of high electron density over each phenylene group of the upper Pc ligand at negative bias voltage. This appearance is consistent with the 16 antinodes of the $\pi$-electron cloud on the upper Pc of the highest occupied molecular orbital of an isolated LuPc$\_2$ molecule, which were calculated with DFT. The Pc ligands are bended so that the planes of the phenylene groups are tilted relative to the planar Pc, as shown in the molecular structure in Figure~\ref{fig:subML}a. Therefore, the outer protrusions appear brighter than the inner ones. Furthermore, we observed a slightly different height of the neighbouring inner as well as outer protrusions of high electron density in the phthalocyanine (refer to the arrows in Figure~\ref{fig:struc}b). This can be explained by a very small distortion of the geometry of the Pc ligand like a twist of the phenylene groups due to the repulsive interaction with the phenylene group of a neighbouring molecule. In reports on the appearance of rare-earth bisphthalocyanines adsorbed on metal surfaces in STM images, only eight bright spots on the phenylene groups around a large central depression were previously observed.\cite{Komeda2011,Zhang2009a}\newline
\indent The structure derived from STM images was validated with molecular mechanics simulations by changing the angle of rotation $\alpha$ of the molecules relative to the unit cell vector and the angle of the unit cell $\varphi$ at fixed intermolecular distances of $A = 1.48$\,nm and $B = 1.54$\,nm. The result is shown in Figure~\ref{fig:LuPc2_uff}. The model for this calculation consists of an arrangement of four LuPc$\_2$ molecules on a supporting graphene sheet. The vertical distance of the outermost carbon atoms of the lower Pc ligand from the graphene was optimised with UFF to be 0.37\,nm in the minimum, which gives an average adsorption energy of $-3\,$eV per molecule. The color scale in the presented plot shows the binding energy for one LuPc$\_2$ molecule in the molecular structure on graphene, which, in particular, corresponds to the total energy of the arrangement subtracted by the energies of the isolated molecules and the adsorption energy on graphene for the other three molecules. In Figure~\ref{fig:LuPc2_uff}a, very high energy barriers can be seen, which result from the strong steric repulsion when the hydrogen atoms are too close to the hydrogen of the neighbouring molecules due to the small distance between the molecules. As shown in Figure~\ref{fig:LuPc2_uff}b, the energetic minimum for the molecular structure is at $\alpha = 67.5^{\circ}$ and $\varphi = 88.25^{\circ}$. We can understand this result when we look at the geometry of a bisphthalocyanine molecule. Because the Pc ligands are rotated by $45^{\circ}$ to each other, if the angle of the molecular axis of one Pc ligand relative to a lattice vector of a square unit cell is $45^{\circ}$, then for the other ligand it would be $0^{\circ}$, but this is unfavourable in close-packed structures because of the steric repulsion. Therefore, the molecules rotate in a way so that the distance of all phenylene groups of the Pc ligands to the ones of neighbouring molecules is the same, which is the case at $\alpha = 22.5^{\circ}$ ($67.5^{\circ}$). Furthermore, the angle of the unit cell is not exactly $90^{\circ}$ because the length of the unit cell vectors was chosen according to the values measured in STM, whereby one unit cell vector is slightly shorter than the other. This is energetically favourable as the length of the lattice vector of the molecular layer, which is parallel to the lattice vector of graphite, becomes an integer multiple of this one, and thus molecules in that direction are adsorbed on equivalent positions of the graphite lattice. However, despite the hexagonal substrate lattice of graphite, a pseudo-square unit cell is preferred for the molecular structure of LuPc$\_2$. From this, it can be concluded that the formation of the energetically most favourable molecular structure is dominated by the intermolecular van der Waals forces with a minor influence of the substrate.
\begin{figure*}[t]
\centering
\includegraphics[width=1.0\textwidth]{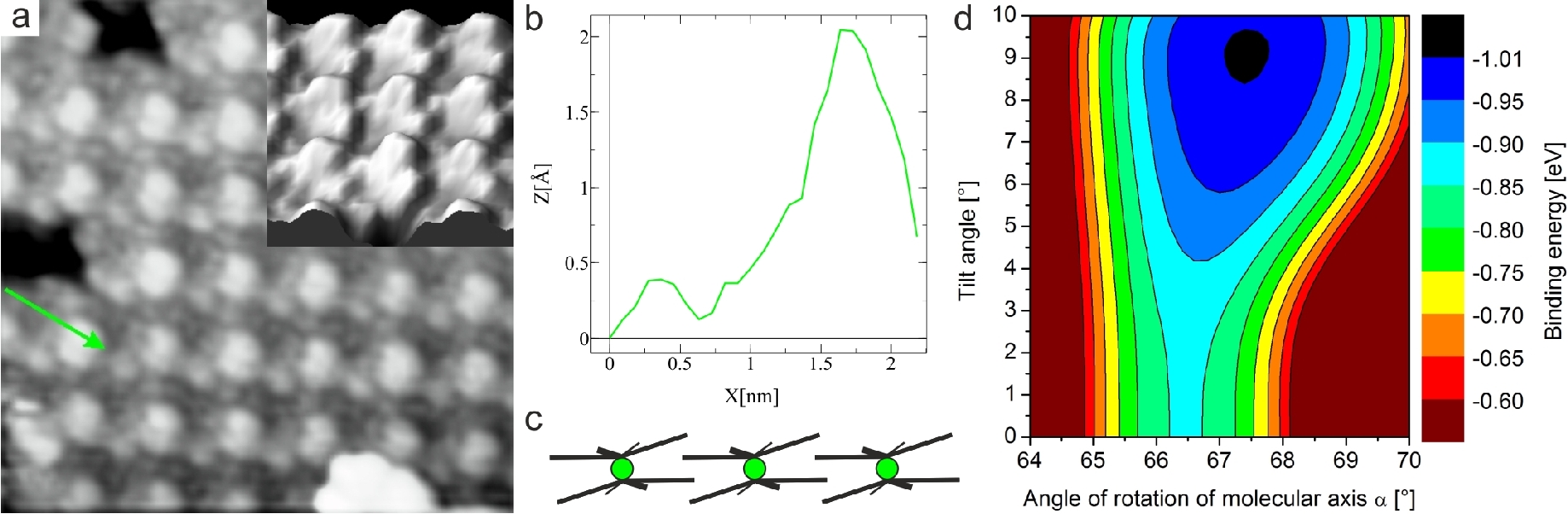}
\caption{(a) STM image ($9.8\times 8.0\,$nm$^2$, $V = 0.6\,$V) of the third molecular layer of LuPc$\_2$ on HOPG with a tilt of $7^{\circ}$ of the molecular plane relative to the plane of the layer. The inset shows a 3D view of the tilted molecules ($5.0\times 5.0\,$nm$^2$). (b) Profile line over one molecule along the green arrow in panel a. (c) Sketch of the arrangement of tilted molecules. (d) Potential energy surface of the tilt angle and rotation of LuPc$\_2$ molecules in the corners of a square unit cell.}
\label{fig:neigung}
\end{figure*}
\begin{figure*}[t]
\centering
\includegraphics[width=1.0\textwidth]{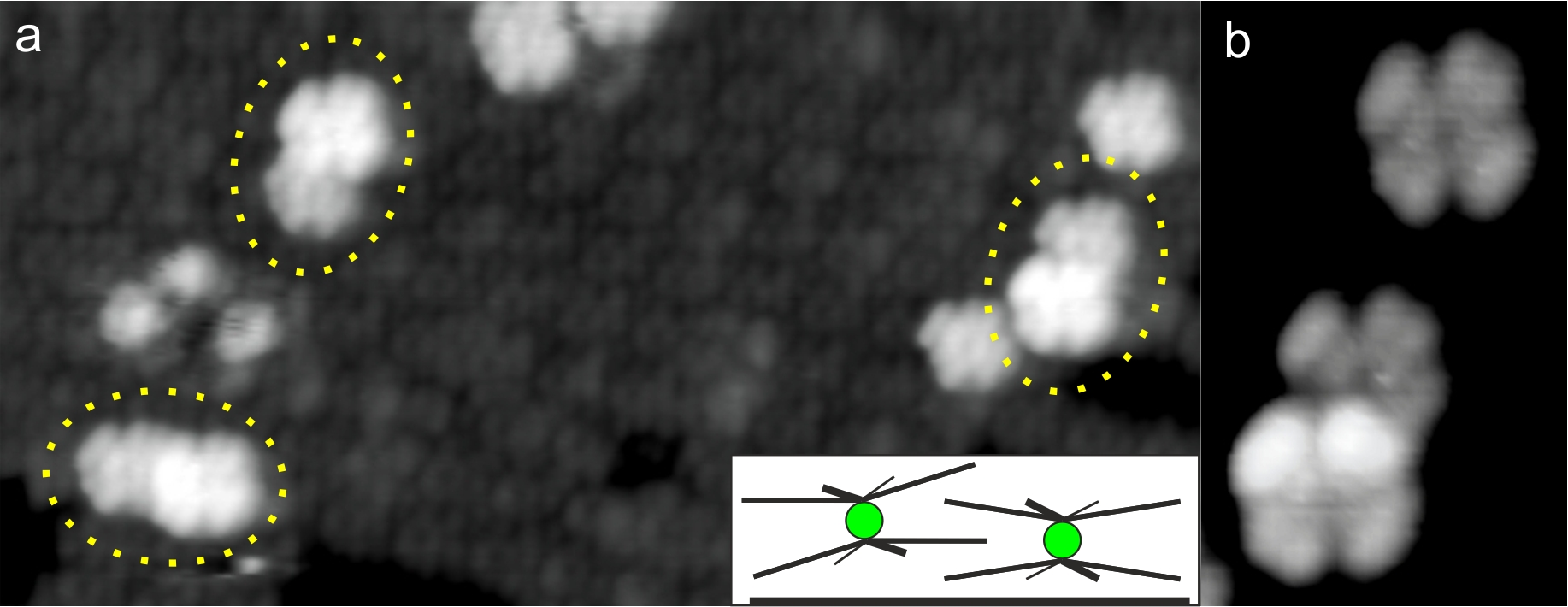}
\caption{(a) STM image ($29.7\times 14.6\,$nm$^2$, $V = -1.5\,$V) of a third ML of LuPc$\_2$ on HOPG with adsorbed single LuPc$\_2$ and three formations of two LuPc$\_2$ molecules with overlapping phthalocyanine ligands (dashed ellipses). The insert shows a sketch of the arrangement. (b) Magnification of an overlapping LuPc$\_2$ pair.}
\label{fig:dimer}
\hrulefill
\end{figure*}
\newline
\indent At multilayer coverage, LuPc$\_2$ shows Volmer-Weber island growth, whereby a molecule in a multilayer is adsorbed on top of the molecule of the underlying layer with a slight horizontal shift. The upper Pc ligand of the adsorbed molecule was found to be rotated by $45^{\circ}$ with respect to the upper Pc ligand of the molecules in the underlying layer so that the adjoining Pc ligands have the same orientation (see Figure~S2 in the Supporting Information). This results in a similar $\pi$-$\pi$ alpha($+$)-stacking as it was found for planar metallophthalocyanines.\cite{Chen2008} The lattice parameters of the molecular structure are the same in the first ML as in the following layers. However, in multilayers, a slight tilt of the molecules relative to the plane of the molecular layer is observed (Figure~\ref{fig:neigung}). Because the tilt of the molecule can be seen at different applied voltages and the distribution of the electron density of the upper Pc ligand has a high symmetry for all molecular orbitals near the Fermi energy, this tilt is a geometric and not an electronic effect. For molecules in the second ML, the tilt angle has been measured to be $(3\pm 1)^{\circ}$ (Figure~S3 in the Supporting Information), and for molecules in the third ML, it has been measured to be $(7\pm 1)^{\circ}$. To comprehend this behavior, in Figure~\ref{fig:neigung}d, we have calculated the binding energy of a LuPc$\_2$ molecule in a molecular layer with the UFF for different tilt angles and with fixed intermolecular distances $A = B = 1.5\,$nm and $\varphi = 90^{\circ}$. The calculation does not include a supporting substrate that would hinder a tilt of the molecules. Therefore, this model is an approximation for the relaxed structure of the top layer in a thick multilayer. The energetic minimum at the tilt angle of $9^{\circ}$ agrees well with our STM results of the third molecular layer. We can assume that the tilt angle of the molecule will still increase slightly in further layers until the minimal energy of the structure is reached. The energetic gain of the minimum energy structure with tilted molecules is $-0.15\,$eV/molecule compared with the structure of a layer where the molecules are not tilted. A small fraction of molecules adsorbed on top of an ordered layer were found to form pairs along one direction of the lattice vectors of the underlying layer (see Figure~\ref{fig:dimer}). The distance between the two molecules in a pair is a little smaller than the intermolecular distance in the ordered layer. Furthermore, the centre of one LuPc$\_2$ molecule is $60\,$pm higher than that of the other LuPc$\_2$, and the upper Pc ligand is slightly tilted and clearly overlaps the lower molecule. From this observation, it is concluded that these molecules intertwine with one phthalocyanine ligand in between the space between both Pc ligands of the other LuPc$\_2$. This leads to a lower repulsion of the adjacent phenylene groups similar to the tilted molecules in the closed layer.
\begin{figure}[t]
\centering
\includegraphics[width=0.49\textwidth]{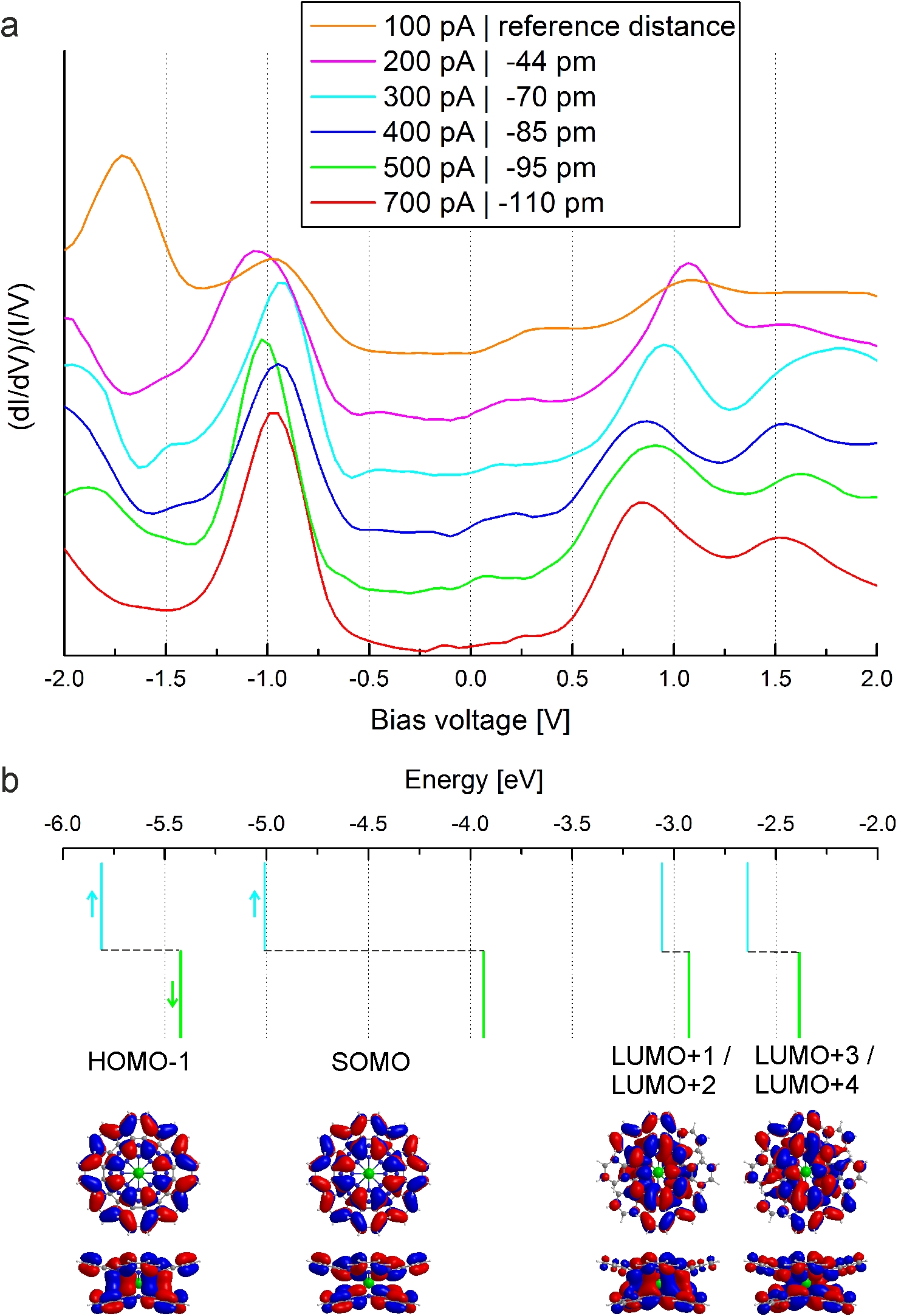}
\caption{(a) Normalized differential conductance spectra of LuPc$\_2$ molecules in a molecular bilayer for different tip-sample distances at a set point bias voltage of $V\_{SP} = 1.2\,$V. A different constant was added to each curve set to distinguish them better. (b) Single-electron energies for the alpha (cyan lines) and beta spin orbitals (green lines) and the corresponding spatial distribution of the electron density for the Kohn-Sham orbitals, calculated with DFT for a geometry-optimised isolated LuPc$\_2$ molecule.}
\label{fig:STS_DFT}
\hrulefill
\end{figure}
\newline
\indent The electronic structure of LuPc$\_2$ on HOPG and the tip$-$sample distance dependence on the orbital-mediated tunnelling has been investigated with tunnelling spectroscopy. Observed peaks in the normalised differential conductance spectra of LuPc$\_2$ molecules in a bilayer adsorbed on HOPG (Figure~\ref{fig:STS_DFT}a) can be assigned to the resonant tunnelling into and out of molecular orbitals. The tip$-$sample separation relative to the distance at the set point current of 100\,pA was obtained for each set point current from $I(z)$-spectroscopy. To understand the electronic structure of LuPc$\_2$, we calculated spin-polarised single-electron energy levels for an isolated LuPc$\_2$ molecule using DFT. The results are shown in Figure~\ref{fig:STS_DFT}b. Because the neutral LuPc$\_2$ molecule consists of two Pc$^{2-}$ ligands and one Lu$^{3+}$ ion, it has an unpaired electron, which is found to be in a $\pi$ orbital localised over both ligands. The different energies for the alpha and beta spin states are obtained because an unrestricted calculation is used for an open-shell electronic system. The singly occupied molecular orbital (SOMO) shows a large energy splitting of 1.07\,eV between the single-electron orbitals at the mPW1PW91/SDD level. The difference between the energy of the occupied SOMO state and the average energy of the alpha and beta spin level of the lowest unoccupied molecular orbital (LUMO) is 2.0\,eV. This is in very good agreement with the STS data for the molecular bilayer of LuPc$\_2$ on HOPG at a set point current of 100\,pA, which means at the largest investigated tip$-$sample distance and thus smallest tip influence. Therefore, the peaks at $-0.95\,$eV and 0.35\,eV might be due to the occupied and unoccupied single-electron states of the SOMO, respectively. The electron affinity level at 1.07\,eV is then assigned to the two-fold degenerated state named LUMO/LUMO$+1$. Vitali \textit{et al.}\cite{Vitali2008} found for TbPc$\_2$ on Cu(111) in the $\text{d}I/\text{d}V$ spectra also a small shoulder at 0.3\,eV, for which they discussed that it arises from the Tb-4f orbitals. In the case of LuPc$\_2$, the f orbitals of Lu are completely filled. Our DFT calculation shows that the lutetium atom has no significant contribution to the molecular orbitals near the Fermi energy, as can be seen in Figure~\ref{fig:STS_DFT}b. This finding is consistent with the STM experiment, where a depression in the centre of the molecule is observed. With decreasing tip$-$sample distance, the energy position of the occupied SOMO level remains pinned besides small fluctuations, and the second ionisation level at $-1.7\,$eV below the Fermi energy, which is assigned to the HOMO$-1$, shifts to about $-2\,$eV. Furthermore, a shift of the unoccupied SOMO state and LUMO/LUMO+1 toward the Fermi energy at 0\,V is observed. It has been reported for a few metallo\-phthalocyanine systems that with a decreasing tip$-$sample distance a shift of the highest occupied or lowest unoccupied molecular orbital relative to the Fermi energy occurs,\cite{Gopakumar2008a} whereas for other systems the peak positions remain fixed.\cite{Deng2003} Shifts in the energy of the molecular orbitals can be due to a charge accumulation in the molecule when at small tip-molecule separations the tunnelling rate from tip to molecule exceeds that from molecule to substrate. A partial filling of the electron affinity levels leads to a shift of the corresponding peaks in STS toward the Fermi energy. An additional electron on the molecule can also have an influence on the energy levels of the other molecular orbitals. Furthermore, the local electric field at the molecule can change depending on the distance to the tip, and at small separations van der Waals forces between tip and molecule atoms could lead to a distortion of the molecule and thus influence the electronic structure.\cite{Kusunoki2001} For bisphthalocyanines, the influence of the tip can also cause a rotation of the upper Pc ligand at small tip$-$molecule distance, which was reported to result in a change of the electronic structure for TbPc$\_2$ on Au(111).\cite{Komeda2011} Thereby, the Kondo peak from the spin of the unpaired electron on the ligand disappeared as the singly occupied molecular orbital became completely filled.

\section{Conclusions}

LuPc$\_2$ molecules adsorbed on HOPG form in self-assembly a highly ordered close-packed molecular structure with an angle of the unit cell of $(88\pm 1)^{\circ}$ and all molecules aligned with the same orientation. In the first layer, the molecules are adsorbed parallel to the surface to increase the $\pi-\pi$ interaction of the Pc ligand with the graphite surface. With subsequent layers, an increasing tilt of the molecules relative to the plane of the layer was observed. This is due to an energy minimisation of the molecular structure when the phenylene groups of the Pc ligand are arranged in between both Pc ligands of the surrounding molecules. For this system, the intermolecular interaction dominates the formation of the structure, whereas the hexagonal graphite substrate defines the direction of one lattice vector of the molecular layer and influences its length. The highly resolved STM images show the distribution of the electron density within a molecular orbital of LuPc$\_2$ in detail. Our STM results and STS at large tip - sample distance, and therefore small tip influence, are in good agreement with DFT calculations for an isolated LuPc$\_2$ molecule and force field structure simulations. This shows that the adsorption on the HOPG substrate has only a small influence on the arrangement and electronic structure of the molecules. These results give insight into the microscopic behavior of rare-earth bisphthalocyanines, which can help to improve the performance of electronic devices with highly ordered molecular thin films and also advance the development of sophisticated molecular spintronic devices.

\section*{Supporting Information}

Overview STM image at multilayer coverage, model for stacking of LuPc$\_2$ molecules, tilt of molecules in the second ML, individual Lu$\_2$Pc$\_2$ adsorbed on an ordered ML of LuPc$\_2$, and xyz coordinates of a geometry-optimised LuPc$\_2$ molecule (DFT).

\section*{Acknowledgements}

This work has been partially financially supported by the Deutsche Forschungsgemeinschaft (DFG) through the research unit 1154 ``Towards Molecular Spintronics''. Computational resources were provided by the ``Chemnitzer Hochleistungs-Linux-Cluster'' (CHiC) at the Technische Universit\"{a}t Chemnitz. We thank Marius Toader for helpful discussions.

\section*{References}

\bibliographystyle{model1a-num-names}
\bibliography{LuPc2_HOPG}

\end{document}